\documentclass[twocolumn,prl]{revtex4}
\usepackage{color}
\usepackage{colordvi}
\usepackage{amsmath}
\usepackage{amssymb}
\usepackage{fancybox}
\usepackage{epsfig}

\begin{document}

\title{Disorder-Induced Stabilization of the Pseudogap in Strongly Correlated 
Systems}

\author{Simone Chiesa}\email{chiesa@ucdavis.edu}
\affiliation{ Dept.~of Physics, University of California,
Davis, CA 95616}
\author{Prabuddha B. Chakraborty}
\affiliation{ Dept.~of Physics, University of California,
Davis, CA 95616}
\author{Warren E. Pickett}
\affiliation{ Dept.~of Physics, University of California,
Davis, CA 95616}
\author{Richard T. Scalettar}
\affiliation{ Dept.~of Physics, University of California,
Davis, CA 95616}

\begin{abstract}
The interplay of strong interaction and strong disorder, as contained in the 
Anderson-Hubbard model, is addressed using two non-perturbative numerical methods:
the Lanczos algorithm in the grand canonical
ensemble at zero temperature and Quantum Monte Carlo.
We find distinctive evidence for a zero-energy anomaly
which is robust upon variation of doping, disorder and interaction
strength. Its similarities to, and differences from,
pseudogap formation in other contexts, including
perturbative treatments of interactions and disorder,
classical theories of localized
charges, and in the clean Hubbard model, are discussed.
\end{abstract}

\maketitle

``Pseudogap" anomalies in the single particle density of states are a
central feature of seemingly disparate materials and models.  On the one
hand, they have an early history in the metal-insulator transition and
the study of the interplay between disorder and inter-particle
interaction.  In the metallic limit of weak coupling and weak disorder,
Altshuler and Aronov (AA) showed\cite{altshuler79}, by means of perturbation
theory, that there is a depression of the spectral density at the
chemical potential, the magnitude of the depression being dependent on
the interaction strength.  In the opposite limit of completely localized
charges (where the model becomes classical),
Efros and Shklovskii (ES)\cite{efros75} have shown that the combined effect
of the unscreened Coulomb potential and disorder also gives
rise to an anomaly at the chemical potential - the Coulomb gap.

On the other hand, in contrast to this situation in which randomness and
electron-electron correlation both are crucial, pseudogap anomalies also
arise in cuprate superconductors and the Hubbard model with {\it no
disorder}.  While particle density does not play a central role in
either the AA and ES pseudogaps, in the high $T_c$ materials the
pseudogap 
is confined to a low-doping region between the superconducting
dome and the parent antiferromagnetic material.  Likewise, in numerical
studies of the Hubbard Hamiltonian \cite{moukouri00,kyung04,kyung06} the
pseudogap is absent for particle densities $\rho < 0.80$.

That disorder is essential to the pseudogap in one situation, yet
present in the clean system in another, raises a fundamental question:
What role does randomness play in low energy anomalies in the density of
states of strongly correlated systems?  In this manuscript, we
suggest that {\em randomness stabilizes the pseudogap}.  Indeed, we
demonstrate two remarkable features of the pseudogap in the disordered
Hubbard model.  First, the density of states anomaly persists in the
limit of an infinitely repulsive local potential $U$, even though the
magnetic energy scale $J \propto t^2/U$ is driven to zero.  Second, it
is independent of doping for a wide range of disorder and interaction
strengths.   The insensitivity of the pseudogap to doping is a novel
effect related to the presence of disorder that has been observed
experimentally\cite{naqib05}.

We consider 
the Anderson-Hubbard Hamiltonian,
\begin{equation}
H(\{\epsilon_i\})=-t\sum_{ij\sigma}\! '
c_{i\sigma}^\dagger c_{j\sigma} +
\sum_{i \sigma}\epsilon_i n_{i\sigma}+|U|\sum_i n_{i\uparrow}n_{i\downarrow},
\label{Hamiltonian}
\end{equation}
where the on-site energies $\epsilon_i$ are sampled uniformly from the
interval $[-\Delta/2:\Delta/2]$, $c^{\dagger}_{i\sigma}\,(c_{i\sigma})$ are
fermion creation(annihilation) operators for site $i$ and spin $\sigma$,
and $n_{i\sigma}=c_{i\sigma}^\dagger c_{i\sigma}$. The primed summation
is on nearest neighbors only.
We diagonalize this Hamiltonian on 10-site square clusters 
using the Lanczos algorithm at $T=0$ and on larger, 64-site clusters using
finite temperature determinant Quantum Monte Carlo (DQMC). As usual,
physical properties are computed as averages over many disorder realizations.

Since both computational methods are well described in a number of
previous publications \cite{dqmc,Lanczos} 
we focus our technical discussion only on the use of the
grand-canonical ensemble to carry out the averaging process.
In the DQMC case this is accomplished naturally as this
method works, by construction, directly within a grand-canonical scheme.
The Lanczos method, on the other hand, is a canonical technique that
diagonalizes the Hamiltonian in sectors of the Hilbert space with 
constant number of particles. Since we are interested in ground state
properties, a given choice of chemical potential 
$\mu$ has the effect of singling out that particle number sector whose 
ground state minimizes $\langle H-\mu N\rangle$. 
Which particular sector is selected is,
of course, dependent on the disorder realization defining
$H(\{\epsilon_i\})$ and
it is not known {\em a priori}. 
Although this requires diagonalization of all particle sectors it also
allows the treatment of non-commensurate fillings.
Since the size
of the system is fairly small we take advantage of the possibility of
changing the boundary conditions of the electronic wave function to
reduce finite size errors.  
The second quantized Hamiltonian becomes therefore a function of $k$, a
vector belonging to the first Brillouin zone of the simulation lattice.
Within the grand-canonical ensemble $k$ is treated as a quenched
disorder variable and sampled uniformly.
This is found to be important in the calculation of the spectral density
for the model considered here. In the limit of vanishing disorder this
scheme reduces to the integration over boundary condition
technique\cite{loh92,gros96}.  With no disorder and no interaction one
recovers the exact spectral density {\it in the thermodynamic
limit}.  

We argue that the density of states obtained on the cluster sizes
used here is relevant both because of the restoration of the
thermodynamic limit by boundary condition averaging, and because
finite size errors become increasingly small in the strong disorder
case on which we focus.  It is also worth noting that, unlike studies of
other phenomena like long range magnetic or superconducting order in the
Hubbard Hamiltonian, which critically rely on finite size scaling on
large lattices, the density of states is less sensitive to system size.
The AA calculation provides a particularly clear example of this statement
as the zero-energy anomaly was predicted using
a class of diagrams that failed to capture localization.

\begin{figure}
\epsfig{file=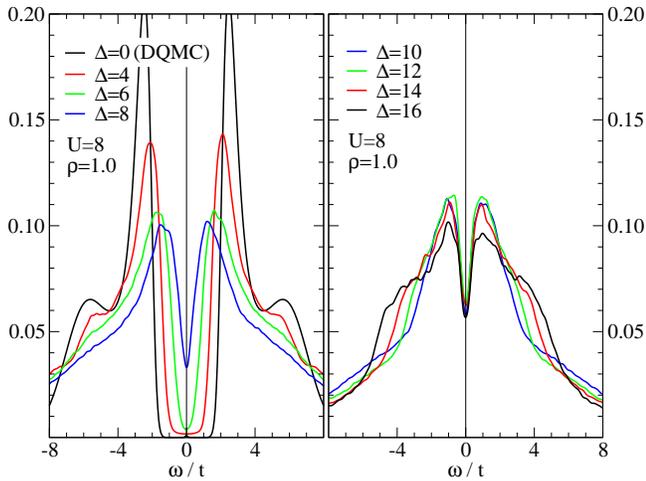,width=0.37\textwidth,angle=-90}
\caption{Evolution of the local spectral density, $A(\omega)$,
at $\rho=1$ for $U=8$ as a function of $\Delta$.  
\underbar{Left panel:} $\Delta \le U$. The Mott gap evident for 
$\Delta=0$ and $4$ is partially filled in as the randomness increases, so that the
value at the Fermi surface, $A(\omega=0)$, becomes finite at $\Delta=U$.
\underbar{Right panel:} $\Delta>U$. A pseudogap survives even for 
large randomness. The spectra are the average of 1000 independent disorder 
realizations.}
\label{halff}
\end{figure}

The disorder-averaged spectral density is given by
\begin{equation}
A(\omega)=\frac{1}{N_s}\sum_{i=1}^{N_s} \frac{1}{\Delta\Omega_\text{BZ}}
\int dk \, d\epsilon \, A_i(k,\{\epsilon_i\},\omega)
\end{equation}
where $N_s$ and $\Omega_\text{BZ}$ are, respectively, the number of sites and
the volume of the Brillouin zone and 
the fermion addition part of $A_i(k,\{\epsilon_i\},\omega)$ is defined as
\begin{equation}
A_i(k,\{\epsilon_i\},\omega)=-\frac{1}{\pi}\text{Im} \left\langle c_i \, 
\frac{1}{\omega-H+E_0+i\eta} \,
c_i^{\dagger}\right\rangle
\label{Aomega}
\end{equation}
and computed using the continued fraction algorithm of Haydock {\em et
al.}\cite{haydock75}. The electron subtraction
spectrum is obtained analogously by interchanging $c_i$ and $c_i^\dagger$. 
In Eq.~\ref{Aomega} $E_0$ is the ground state energy of
$H\equiv H(k,\{\epsilon_i\})$ and $\eta$ is a small real parameter giving the
broadening of the $\delta$ functions constituting the spectrum.
$A(\omega)$ is directly measured in tunneling spectroscopy,
photoemission and inverse photoemission experiments. Since disorder averaging
restores particle-hole symmetry, it suffices to
consider the evolution of $A(\omega)$ at densities $\rho\le 1$.
In DQMC, $A(\omega)$ is obtained by a maximum entropy analytic
continuation of the imaginary time Green's function \cite{gubernatis91} 
computed using periodic boundary conditions ($k=0$).

\begin{figure}
\epsfig{file=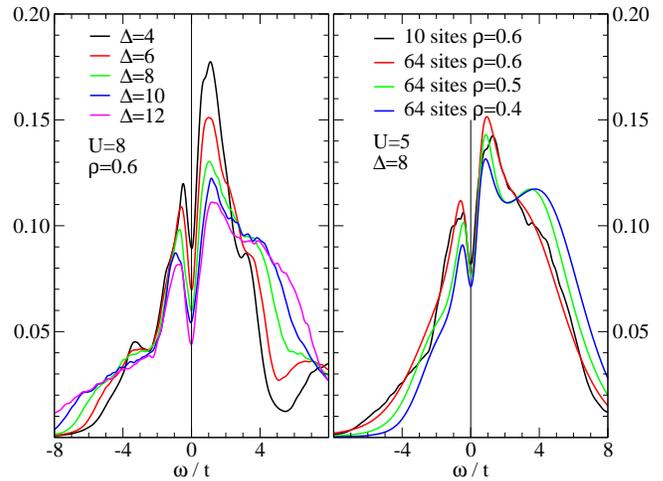,width=0.37\textwidth,angle=-90}
\caption{Evolution of the local spectral density away from half-filling 
(no Mott gap at $\Delta=0$).  
\underbar{Left panel:} $A(\omega)$ as a function of $\Delta$.
$\Delta$ ranges from well below $U$ to well above.  The
dip in $A(\omega)$ at $\omega=0$ becomes deeper as $\Delta$ increases.
In the clean limit $\Delta=0$ we observe
no pseudogap this far from half-filling, in agreement with
[\onlinecite{moukouri00,kyung04,kyung06}].  
\underbar{Right panel:} DQMC results on a 8x8=64 site lattice at $T=t/6$
averaging over 64 disorder realizations.
The scale (width) of the pseudogap is largely independent of doping.
For $\rho=0.6$ the 10-site Lanczos result is also reported. Despite the
different cluster size the agreement between the two techniques is excellent.
}
\label{doped}
\end{figure}

There are three different independent parameters ($t$ sets the unit
of energy) that we examine when analyzing the spectral densitiy:
the interaction $U$, the disorder $\Delta$ and the doping
$\rho$.  We start by considering the dependence on $\Delta$ for the
half-filled case and constant $U=8$.  
On increasing $\Delta$,
for $\Delta<U$, the Mott gap present at $\Delta=0$ is gradually
filled in, and the evolution of the spectral density follows the expected
trend (see left panel of Fig.~\ref{halff}).  For $\Delta>U$ one might
expect the residual dip to disappear completely, at least at large
enough disorder, but this is not what is found numerically. The right
panel of Fig.~\ref{halff} shows the  behavior for $\Delta=10,12,14,16$:
Increasing the disorder above $\Delta=U$ leaves a residual pseudogap
independent of the disorder strength. 
There is a sharpness in the behavior of $A(\omega)$ at
small $\omega$ which suggests that the anomaly could be
non-analytic at $T=0$. However, finite size rounding 
prevent a precise characterization of this feature.

The appearance of a pseudogap is remarkably different from what is found
using dynamical mean field theory (DMFT) in a similar range
of parameters\cite{song08} for the same Hamiltonian. 
DMFT predicts, at least in the Hubbard-I approximation\cite{song08}, 
that $A(\omega)$ evolves smoothly as the energy crosses the chemical potential,
with no pseudogap.
This indicates that the anomaly is likely to be determined by non-local, 
short-ranged correlations and connected to reduced dimensionality.

Let us now move to the incommensurate filling $\rho=0.6$, where there is
no Mott gap.  $A(\omega)$ is plotted in Fig.~\ref{doped}. A
pseudogap is now evident at all disorder strengths. For $\Delta>U$ it
behaves as for the $\rho=1$ case and saturates for strong enough disorder. In
the $\Delta<U$ regime the pseudogap gets deeper with increasing disorder
but its width remains largely unchanged. 
In contrast, the depth of the pseudogap which develops out of the Mott
phase at $\rho=1$ decreases as $\Delta$ increases\footnote{Eventually,
in the limit $\Delta\rightarrow\infty$, the density of state has
to tend uniformly to $0$ because of the normalization constraint so that 
this trend is inverted.}.
Such a difference is not surprising since, at $\rho=1$, there is a
cross-over from a $U>\Delta$ regime, dominated by the Mott-gap scale to
the $\Delta>U$ regime characterized by the pseudogap scale. On the other
hand, away from half filling, the pseudogap scale is always the leading
instability and continues to persist in the highly disordered regime. It
remains unclear why the depth of the pseudogap saturates instead
of moving monotonically to $0$ as $\Delta$ increases.
Certainly the behavior of the anomaly away from
half-filling makes evident that the naive picture in which the effect of 
disorder is only to smear the spectral density is inappropriate.

\begin{figure}
\epsfig{file=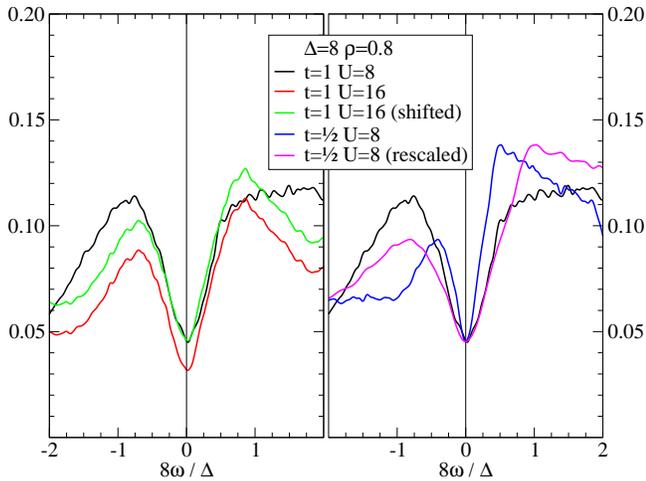,width=0.37\textwidth,angle=-90}
\caption{Scaling of the pseudogap as a function of interaction and hopping.
The key observation is that the pseudogap is unchanged when $U$
increases from $U=8$ to $U=16$ (left panel), and hence the antiferromagnetic exchange 
$J$ is halved, but has a width which is proportional to $t$ (right panel). The shifting
in the third data set is done so as to make the two minima, for $U=8$ and
$U=16$, coincide. In the last
data set the energy axis was rescaled by a factor of two.
}
\label{scale}
\end{figure}

We have carefully studied the doping dependence of the pseudogap, 
especially since in the clean model it is present only for small
doping.  In contrast,
for $\Delta=10$ and $U=8$ the spectral density shows a
nearly universal pseudogap: $A(\omega)$ 
for $\rho=0.6 - 1.0$ coincide over the entire
pseudogap region. 
This feature persists in the
$\Delta\le U$ regime as soon as the system is sufficiently away from the 
Mott-gap dominated regime. Typically, at $\rho=0.9$, a distinct dip 
at $\omega=0$ is seen at all disorder strengths considered in this work.
At densities $\rho\le0.6$ and $U\le 5$ the sign problem in the DQMC method is 
mild and does not prohibit obtaining accurate spectral functions
\footnote{Strong on-site disorder leads
to a severe sign problem at half-filling contrary to the $\Delta=0$ situation}.
The agreement between the two numerical techniques
at $\rho=0.6$ and $\Delta=8$ (see right panel of Fig.\ref{doped}) is excellent and gives a 
firmer basis to our speculations on the irrelevance
of the cluster size that was exactly diagonalized.
DQMC results from $\rho=0.6$ to $\rho=0.4$ also confirms the stability
of the pseudogap under doping.
That strong disorder can stabilize the pseudogap was also
observed experimentally\cite{naqib05} in a conductivity study of YBCO samples where Cu was
substituted with Zn. 
The $T-\rho$ phase diagram of this heavily disordered system
shows a flat pseudogap crossover line
in agreement with our finding of a doping-independent pseudogap energy scale.

That variation in $\Delta$ or $\rho$ both leave the anomaly unchanged can be
understood as follows. 
Consider, for instance, the half-filled case with $t=0$
and $\Delta>U$ in which a finite fraction of the sites remains empty.
When $t\ne 0$, electrons lower their energy by delocalizing on neighboring
empty sites, regardless of the overall density, as the physics is
primarily local.
The effect of doping does not alter this situation but merely shifts
the ``action''
on the new set of sites lying close to $\mu$ . Although the details of
the effect of $t$ are ultimately responsible for the formation of the pseudogap,
one can see that {\em if} the pseudogap forms and is stable upon variation in
disorder {\em then} it follows that the same anomaly forms and is stable
upon variation of doping.

The distinctive feature of the strongly disordered Anderson-Hubbard
Hamiltonian is that particle localization occurs independently of
the doping level. In the non-disordered case, localization is induced by
correlation and gradually disappears as the density moves away from
half-filling. 
Despite this different doping dependence, there are similarities with
the ordered case that point to a common localization-related origin of
the pseudogap.  
We found that the pseudogap scale is independent of $U$
for very large $U$ with $A(\omega=\mu)>0$ at any finite doping.
To make this point more quantitative we show the
$U=8$ and $U=16$ pseudogaps in the left panel of Fig.~\ref{scale} at the
common disorder strength $\Delta=8$. One can see that, apart from a
deepening of $A(\omega)$ with increasing $U$, the
pseudogap scale is essentially unchanged. 
Independence of the pseudogap from $U$
and $\Delta$ in the strongly disordered and strongly interacting regime
leaves $t$ as the only energy scale relevant to the
pseudogap phenomenon. Indeed, halving $t$ (right panel of Fig.\ref{scale})
induces an almost linear reduction on the energy scale characterizing the anomaly.
This suggests a kinetic mechanism for pseudogap
formation and it certainly rules out super-exchange
since the magnetic energy scale $J$ tends to $0$ in this limit. Spin, however, has 
to play a crucial role since, obviously, spinless fermions interacting through
a local $U$ would show a smooth
$A(\omega)$ with no suppression at the Fermi surface.

\begin{figure}
\epsfig{file=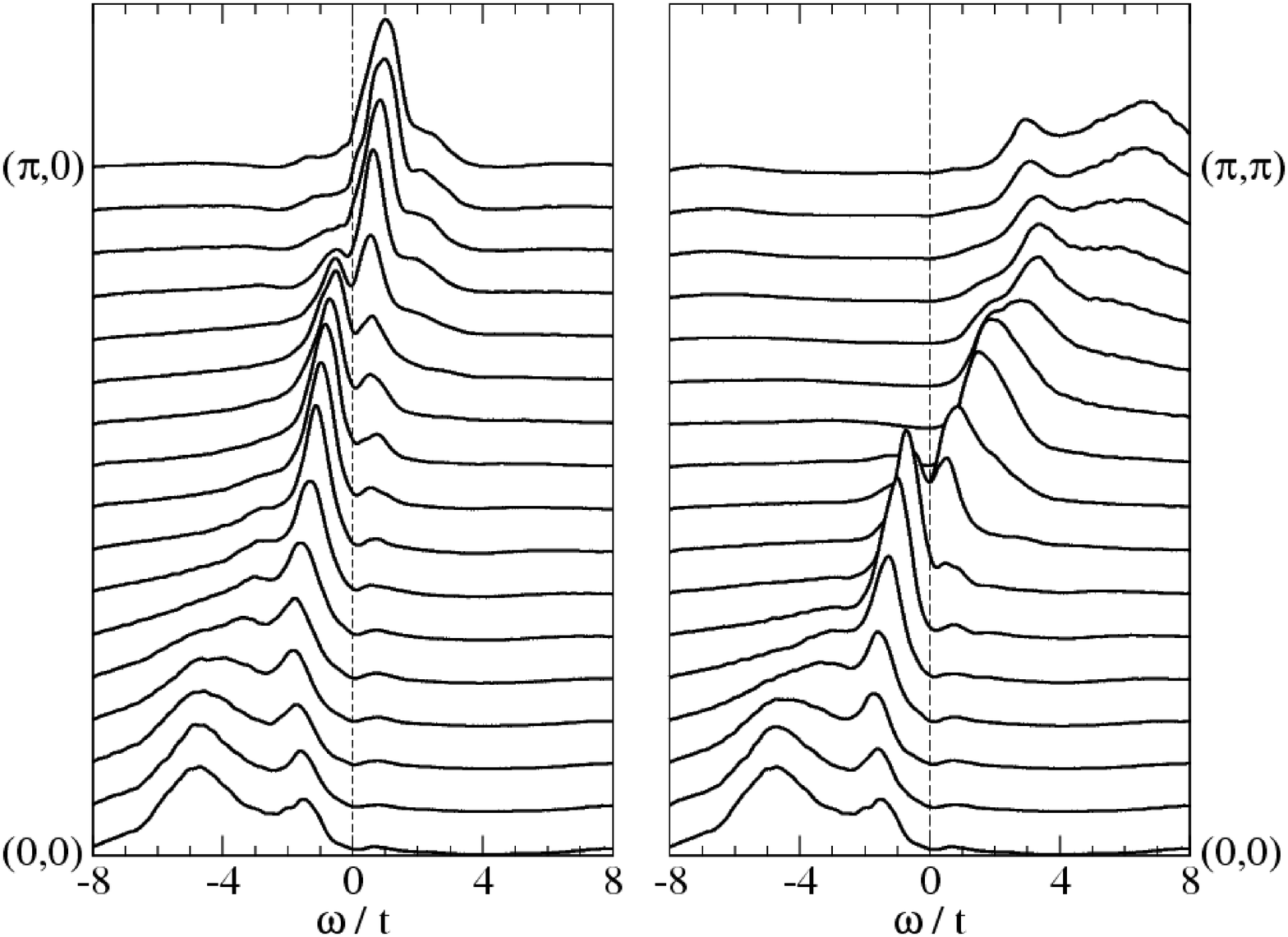,width=0.47\textwidth,angle=0}
\caption{
The momentum resolved spectral function $A_k(\omega)$ exhibits
a pseudogap as the Fermi surface is crossed.
Here $U=8$, $\Delta=8$, and $\rho=0.8$.
}
\label{disperse}
\end{figure}

An interesting feature of the clean Hubbard
model and of experiments on cuprates is the momentum dependence of 
the pseudogap. In the hole doped case for example, sharp excitations 
survive as one crosses the Fermi energy along the nodal direction 
whereas a pseudogap develops along the anti-nodal one.
The behavior of the averaged $A_k(\omega)$ in the strongly disordered
regime ($\Delta=8$) is given in Fig.~\ref{disperse}. 
As $k$ cuts through the Fermi surface either along the anti-nodal ($(0,0)$ to
$(\pi,0)$) or nodal ($(0,0)$ to $(\pi,\pi)$) lines, a depression in 
$A_k(\omega)$ is seen as $k_F$ is traversed. This is therefore at odds
with the results on the clean model and experimental data. A qualitative
explanation for such a difference could reside in the 
different nature of electron localization in the two cases:
while localization is certainly isotropic in the strongly disordered scenario 
this is not necessarily so in clean materials where it is induced 
by dynamical inhomogeneities.

In conclusion, we have shown the formation of a robust pseudogap in
systems with strong repulsive local interaction and strong disorder.
This parameter regime lies outside the range of applicability of the
perturbative AA calculation.  It also lies outside the
Coulomb gap scenario since the potential is local, and there is no
Coulomb gap for the on-site Hubbard interaction when the itinerancy of
the electrons is switched off.

Although our exact Lanczos analysis is carried out only on clusters
small enough to be exactly diagonalized, the phenomenon is also present
on much larger clusters treated with the exact DQMC method.
The pseudogap energy scale is set
by $t$, a result shared with other recent numerical studies of
non-disordered, strongly interacting systems \cite{kyung04,kyung06}.
In particular, it persists even when $J \rightarrow
0$, and so does not appear to be linked to antiferromagnetic
fluctuations.  Finally, it is suggestive that recent experiments 
\cite{naqib05}
observed a doping independent anomalous behavior in the conductivity of
highly disordered cuprates consistent with what is reported here.

In the uniform Hubbard model, and the cuprate materials which it may
describe, spatial inhomogeneities arise spontaneously,
without any explicit symmetry-breaking in the Hamiltonian itself
\cite{mcelroytranquada,white03}.  Likewise, the pseudogap is a feature of the
model \cite{moukouri00,kyung04,kyung06} and the materials.  Both phenomena
disappear with doping. Disorder however induces
i)the stabilization of the pseudogap over a much larger range of densities;
ii) its independence on the  momentum.
These facts, together with the persistence of the pseudogap at large values of $U$
in both disordered and uniform models, are compatible 
with a scenario where the pseudogap in the strong-coupling regime is intrinsically linked to electron localization
and driven by a kinetic mechanism.
Such a picture, which
invokes spatial inhomogeneities, thus connects the appearance of the
pseudogap in the two seemingly rather different contexts of
metal-insulator transitions driven by the interplay of randomness and
correlation, and the cuprate superconductors.

This research was sponsored by the National Nuclear Security
Administration under the Stewardship Science Academic Alliances program
through DOE Research Grant DOE DE-FG01-06NA26204.  We are grateful for
input from J. Garcia, to M. Jarrell for use of his Maximum Entropy
code and to R. Singh and G. Zimanyi for many useful discussions.

\end{document}